**Terahertz-driven Luminescence and Colossal Stark Effect in CdSe:CdS Colloidal Quantum Dots**

**Brandt C. Pein(1*), Wendi Chang (2), Harold Y. Hwang (3), Jennifer Scherer (1), Igor Coropceanu (1), Xiaoguang Zhao (4), Xin Zhang (4), Vladimir Bulović (2), Moungi Bawendi (1) and Keith A. Nelson (1)**

(1) Department of Chemistry and (2) Department of Electrical Engineering and Computer Science, Massachusetts Institute of Technology, Cambridge, MA USA.

(3) Massachusetts Institute of Technology Lincoln Laboratory, Lexington, MA USA.

(4) Department of Mechanical Engineering, Boston University, Boston, MA USA.

*Corresponding author: bpein@mit.edu.

**Abstract**:

Unique optical properties of colloidal semiconductor quantum dots (QDs), arising from quantum mechanical confinement of charge within these structures,(*1*) present a versatile testbed for the study of how high electric fields affect the electronic structure of nanostructured solids. Earlier studies of quasi-DC electric field modulation of QD properties(*2–5*) have been limited by the electrostatic breakdown processes under the high externally applied electric fields, which have restricted the range of modulation of QD properties. In contrast, in the present work we drive CdSe:CdS core:shell QD films with high-field THz-frequency electromagnetic pulses whose duration is only a few picoseconds. Surprisingly, in response to the THz excitation we observe QD luminescence even in the absence of an external charge source. Our experiments show that QD luminescence is associated with a remarkably high and rapid modulation of the QD band-gap, which is changing by more than 0.5 eV (corresponding to 25% of the unperturbed bandgap energy) within the picosecond timeframe of THz field profile. We show that these colossal energy shifts can be consistently explained by the quantum confined Stark effect. Our work demonstrates a route to extreme modulation of material properties without configurational changes in material sets or geometries. Additionally, we expect that this platform can be adapted to a novel compact THz detection scheme where conversion of THz fields (with meV-scale photon energies) to the visible/near-IR band (with eV-scale photon energies) can be achieved at room temperature with high bandwidth and sensitivity.

**Main Text:**

Quantum confinement of charge in QDs is responsible for their tunable magnetic, electronic and optical properties, which have facilitated the development of QD-based photovoltaics, photodetectors, light-emitting devices (LEDs), and other photonic and biomedical imaging technologies. QDs can be actively manipulated with externally applied electric and magnetic fields, providing routes to a new class of multifunctional photonic devices with controllable absorption and emission properties(*6*, *7*). The extent to which those optical and electronic QD properties can be controlled with extremely large external fields is still largely unexplored.

A series of recent studies demonstrated nonequilibrium manipulation of material properties driven by intense, THz-frequency radiation including solid-solid phase transitions(*8*), impact ionization(*9–11*), and field ionization(*12*). In many cases, inherent technological challenges related to achieving high THz field amplitudes have been mitigated by concentrating THz radiation into subwavelength metallic structures such as split ring resonators and antennas(*8*, *12–15*). Such demonstrations have opened up access to multi-MV/cm field strengths for nonlinear THz spectroscopy applications. In the present work, we applied field enhancement techniques similar to those described above to develop a novel platform to study manipulation of QD film optical properties on a picosecond timescale.

To concentrate the incident THz radiation we use a simple field enhancement structure (FES) design, a metal microslit array on a $SiO_2$ substrate, which consists of parallel gold lines with 98 µm widths, separated by 2 µm capacitive gaps. Such an FES offers a route to field enhancement factors intermediate between dipole antennas and nanoslit arrays(*16*, *17*). A THz field polarized perpendicular to the gold lines (Fig. 1a) is directed onto the microslit array (Fig. 1b) and undergoes enhancement of up to 2 orders of magnitude in the capacitive gaps. QDs deposited over the microslit array are therefore subjected to the enhanced THz field.  Two samples were tested: sample I consists of QDs on top of bare gold microslits, and sample II has QDs on top of a 65±15 nm silica layer that electrically insulates the QDs from the gold microslits. (See Supplementary Materials for details of the experimental apparatus and samples.)

Figure 1c shows a visible light image of THz-driven luminescence (THz-L) originating from QDs within the FES gaps of sample II. The inset shows an enlargement of a single gap region, indicating that QD luminescence is enhanced in the vicinity of the FES gaps. Figure 2 shows the spectral and kinetic characteristics of the THz-L along with photoluminescence (PL) induced by $\lambda = 400$ nm wavelength light. The THz-L central wavelength is $\lambda = 620$ nm (2.0 eV; see Fig. 2a). Depending on the incident peak field strength, the THz-L spectra FWHM are broadened by as much as 16 meV and redshifted up to 12 meV relative to the PL spectrum generated by $\lambda = 400$ nm pulses (Fig. 2a). The time-dependent PL from both samples could be fit well to a mono-exponential decay with a 15 ns lifetime (Fig. 2b).  In contrast, the THz-L emission at all incident field strengths exhibits multi-exponential decay kinetics with shorter lifetime components compared to those observed in the PL decay. We ascribe this difference to QD charging under THz excitation. Excitons in charged QDs have additional non-radiative pathways, such as Auger recombination, which would reduce the observed radiative lifetime.(*2*, *18*) The presence of local electric fields due to QD charging would also explain the observed redshift and broadening of the THz-L spectra.

Exciton formation in QD-LEDs driven by DC and quasi-DC electric fields requires long-range transport of electrons and holes. Given the picosecond period of a THz pulse and the low charge carrier mobility of QD films ($10^{-4}$ to 1 cm$^2$/(Vs)), long-range transport of electrons and holes across a microslit gap is unlikely(*19–21*). With our largest incident THz field strength of 320 kV/cm, the peak field across most of the gap region is approximately 2 MV/cm. (As discussed below, the peak field near the edges of the gaps, from where the EL emerges, reaches approximately 15 MV/cm, but charge transport across the gap would depend on the field strength in the center.) With a mobility of 1 cm$^2$/(Vs) and a 2 MV/cm, electric field, in 1 ps the charge could travel a distance of 20 nm, a negligible fraction of the microslit gap width. It is clear that we cannot describe our observations as analogous to DC or quasi-DC driven luminescence. Due to the limitations imposed by the RC delay in a typical electrically-driven capacitive QD-LED structure, earlier a.c.-driven luminescence measurements have been limited to frequencies of up to 1 MHz, which is slow enough to allow charge migration between electrodes to occur.(*22–25*) However, even in the absence of long-range migration, an electrode-driven mechanism for EL is possible(*26–28*) in which the alternate polarities of an a.c. field could inject electrons and holes into the conduction and valence bands (CB and VB) respectively of a QD located next to one electrode, leading to exciton formation and luminescence. Alternatively, a purely field-driven(*22, 25, 29, 30*) mechanism, based on local QD-to-QD charge exchange, could occur. When the voltage drop between neighboring QDs is equal to or greater than the QD band gap, VB electrons of one QD can be emitted through field ionization and tunnel into the CB of a neighboring QD, leaving one QD charged positively and the other negatively. In the next electric field half-cycle the VB electrons of the second QD could be transferred to the neighboring QD, leading to formation of a QD exciton and consequent luminescence. To explore the likelihood of these mechanisms occurring in our structures we next analyze the THz-field-dependent luminescence intensity. The THz-L intensity depends nonlinearly on the peak THz field. Below a threshold field level $F_{th}$ no emission occurs. Above the threshold, the dependence of the THz-L intensity on the incident field level is strongly nonlinear. $F_{th}$ is 85±5 kV/cm and 180±7 kV/cm for samples I and II respectively (Fig. 3a). The THz-L signals near $F_{th}$ are assumed to originate from QDs residing in regions with the strongest field enhancement. We model the spatial variation of the THz electric field in the microslit structures, using Computer Simulation Technology (CST) microwave studio,(*31*) and based on the peak time-dependent field value we calculate the field enhancement factor $\beta$ as a function of the perpendicular lateral direction $x$ (the incident field polarization direction), and the height $z$ above the $SiO_2$ substrate (Fig. 3b). (See Supplementary Materials for details of CST simulation.) For a given height, $z$, above the substrate, the change in electric potential $\Delta V$ at position $x$ from one gold surface toward the other is computed as

(1) $$\Delta V(x, z) = \int_0^x \frac{\beta(x-x',z) F_{THz}}{\varepsilon_{film}} dx'$$

for an incident THz electric field with peak amplitude $F_{THz}$. Since the THz field is oriented along the x-direction we neglect polarization components along the z-direction. We use a dielectric constant $\varepsilon_{film} = 3$ for the QD film, based on the Maxwell-Garnett effective medium approximation from previous work on CdSe:CdS colloidal QD thin films.(*4*, *25*) We also assume that the adjacent center-to-center distance of neighboring QDs is 10 nm, based on TEM measurements (see Supplementary Materials). The largest calculated voltage drop occurs between QD neighbors situated as close as possible to one of the gold electrodes and the oxide substrate, as shown for sample I in Fig. 3d. For these locations for the two samples the voltage drop along the *x*-direction as a function of incident peak THz field is shown in Figure 3c. Assuming the field-driven model of exciton generation (Fig. 3d), $F_{th}$ is reached when the voltage drop between the neighboring QDs equals the QD bandgap energy (2 eV) divided by the electron charge. The simulated $F_{th}$ values are 86 kV/cm and 186 kV/cm for samples I and II respectively, in excellent agreement with the experimentally measured values of 85±5 kV/cm and 180±7 kV/cm, respectively.

We note that in sample I, electrode-driven carrier injection into the QD film is possible in which case the onset of THz-L would occur when carriers begin tunneling into the QD layer from the gold microslits (Fig. 3e). Here we compare our results to the Fowler-Nordheim charge tunneling model used in previous works concerning charge injection into low mobility semiconductors and THz-driven field emission from gold microstructures.(*12*, *32*, *33*) Charges injected into low mobility semiconducting materials localize at the interface and backflow into the source of the carriers due to thermionic emission. This results in a net low interfacial current density and, although the enhanced THz field can be quite strong, the picosecond oscillation period limits the total number of carriers that can be injected per pulse. The Fowler-Nordheim equation for interfacial current density is

(2) $$J_{FN} = a(\frac{F_{THz}}{\varepsilon_{film}}\beta)^2 \varphi^{-1} exp\left(\frac{-\varepsilon_{film} b \varphi^{\frac{3}{2}}}{\beta F_{THz}}\right)$$

where $a = 1.53 \cdot 10^{-6}$ A·eV·V$^{-2}$ and $b = 6.83 \cdot 10^9$ eV$^{-3/2}$·V/m are combinations of universal constants.(*34*) The average field enhancement nearest to the surfaces of the gold microslits residing within the gaps is $\beta = 106$. The potential energy barrier height $\varphi$ is the barrier electrons or holes must overcome to be injected into the QD film. The value of $\varphi$ can take on a range of values depending on the nature of the QD/gold interface. We assume that the Fermi level of the gold microslits lies in the middle of the QD band gap, giving an estimated barrier of 1 eV for electrons and holes. Due to the time-dependent THz electric field asymmetry with respect to the troughs and peaks of the field (Fig. 3e), there are different field dependencies for electrons and holes on both sides of a gap. Thus on one side of a gap, more electrons will be injected than holes and on the opposite side more holes will be injected than electrons. Integrating $J_{FN}$ over the THz field temporal profile and the slit surface area gives the number of electrons or holes injected within a slit gap for a given peak incident THz field (Fig. 3f) (see Supplementary Materials for details on calculation). One electron-hole pair is required to form an exciton, so the threshold incident peak field is predicted to be 219 kV/cm, which is significantly higher than the experimental result for sample I. We therefore conclude that THz-driven luminescence is, at least at lower incident THz field amplitudes, primarily due to field-driven QD-QD interactions in both samples I and II.

At the highest incident THz field amplitudes, regions of the QD layer can experience electric fields as large as 15 MV/cm on the picosecond timescale while the THz pulse is still present. This can have a significant impact on the optical absorption properties due to the quantum-confined Stark effect (QCSE) in which an electric field distorts the QD potential energy surface leading to an effective band gap reduction (Fig. 4a). To probe this electroabsorption effect, femtosecond optical pulses with photon energies between 1.45-1.8 eV (all below the QD bandgap of 2.0 eV) were spatially and temporally overlapped with the THz pulses incident on the microslit structures.

The optical pulses induce no detectable PL by themselves or when they arrive before or after THz pulses. However, when an optical pulse (with either polarization) is time-coincident with a peak of the THz field it induces emission, which is consistent with THz-field-induced electroabsorption. Notably, as the optical pulse is variably delayed across the THz pulse field profile, it generates a PL signal whose intensity approximately tracks the absolute magnitude of the THz field (Fig. 4b). For the data in Figure 4c, we fix the optical excitation pulses at zero delay (relative to the THz field peak) and vary the THz field amplitude to determine the $F_{th}$ value for electroabsorption. The measurement is repeated for different incident optical pump photon energies. As the optical pump photon energy is increased $F_{th}$ is reduced, consistent with THz-field-induced electroabsorption. At each threshold field level, the QD band gap energy is reduced from its field-free value to the optical pump photon energy (Fig. 4d). For a 100 kV/cm incident THz field (near the onset of THz-induced EL) the band-gap is reduced by more than 0.5 eV which is an order of magnitude larger than previously demonstrated Stark effect measurements of QDs.(*3*) At higher incident fields the QD band gap is likely reduced further, however the overlap of electroabsorption-enhanced luminescence with purely THz-driven luminescence signals complicates quantitative determination of the bandgap reduction.

This work demonstrates that ultrafast THz pulses can drive a luminescence response in a colloidal QD film. While electrode-driven charge injection and purely field-driven mechanisms can both contribute to THz-L, our results indicate that the field-driven mechanism dominates even when charge injection is possible through direct contact between the gold lines of our microslit structure and the QD emitters. THz-L is driven by direct distortion of the QD electronic structure, and we observe a modulation of the QD bandgap of more than 0.5 eV, shifting the QD absorption across a significant part of the visible spectrum. The QD emission and spectral shifting driven by the THz field may be used as a novel approach to THz detection and imaging and for ultrahigh-frequency electro-optic modulation.

**Figures:**

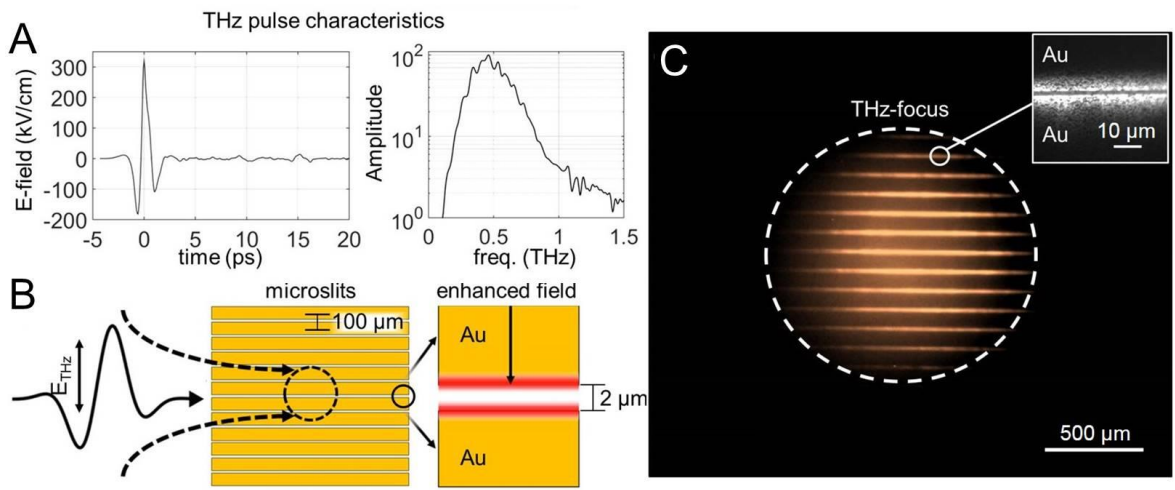

**Fig. 1 THz field-enhancing microslits and QD luminescence.** (**A**) The incident free-space single-cycle THz pulses have a maximum 320 kV/cm electric field and a spectrum centered at 0.5 THz. (**B**) Microslit array consisting of parallel horizontal gold lines on an amorphous silica substrate was irradiated by vertically polarized THz pulses. The THz field was enhanced in the 2-μm capacitive gaps between the gold lines. Quantum dots were deposited over the structure, and the QDs in the gap regions were subjected to the enhanced fields. Samples I and II were fabricated respectively without and with a 65±15 nm oxide layer deposited over the structure prior to the QDs. (**C**) Focusing THz pulses onto either sample I or II generates a visible light image. The image shown is from sample II. The light originates from QD THz-L coming from within the microslit gaps. The inset shows an enlarged image of one gap region, showing that the THz-L is most intense from the regions near the edges of the gaps.

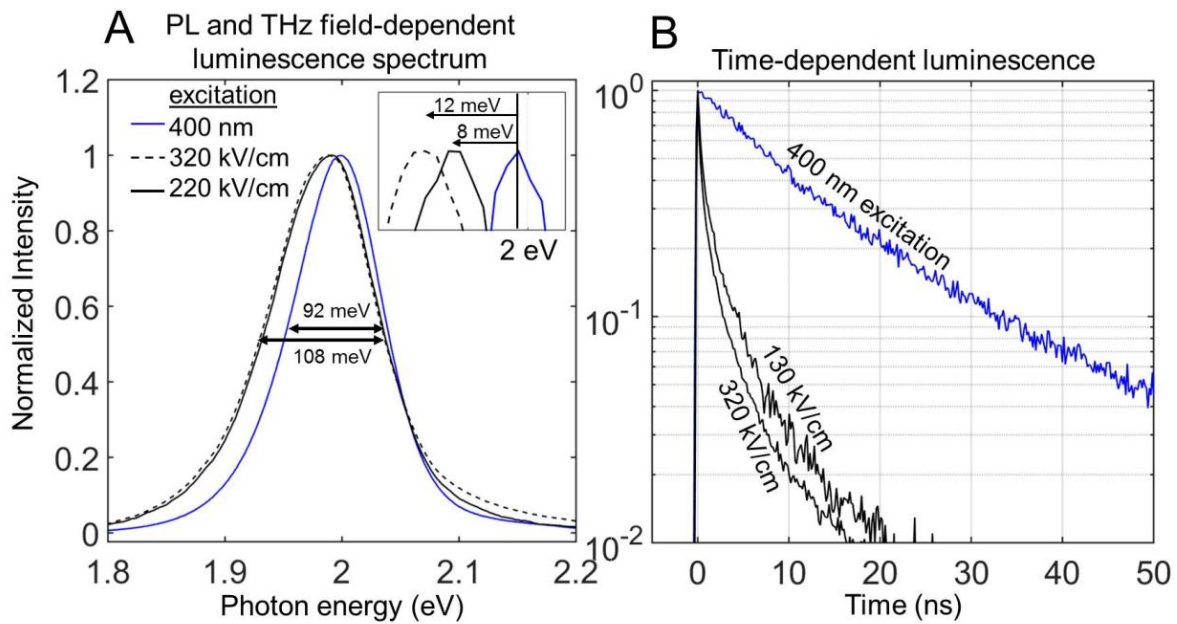

**Fig. 2 Characteristics of THz-induced luminescence and optically induced photoluminescence originating from microslit samples.** (**A**) The PL spectrum (blue curve) generated by λ = 400 nm excitation wavelength is centered at 2.0 eV. The THz-driven luminescence spectrum is broadened relative to the λ = 400 nm excited spectrum. (Inset) The peak wavelength of THz-driven THz-L emission is redshifted by as much as 12 meV at the highest incident THz peak field compared to PL emission. (**B**) Following excitation of the sample with λ = 400 nm pulses, the time-dependent photoluminescence follows a single exponential decay with a 15 ns lifetime. Excitation with 320 kV/cm or 130 kV/cm peak incident THz field generates luminescence with a faster multi-exponential decay.

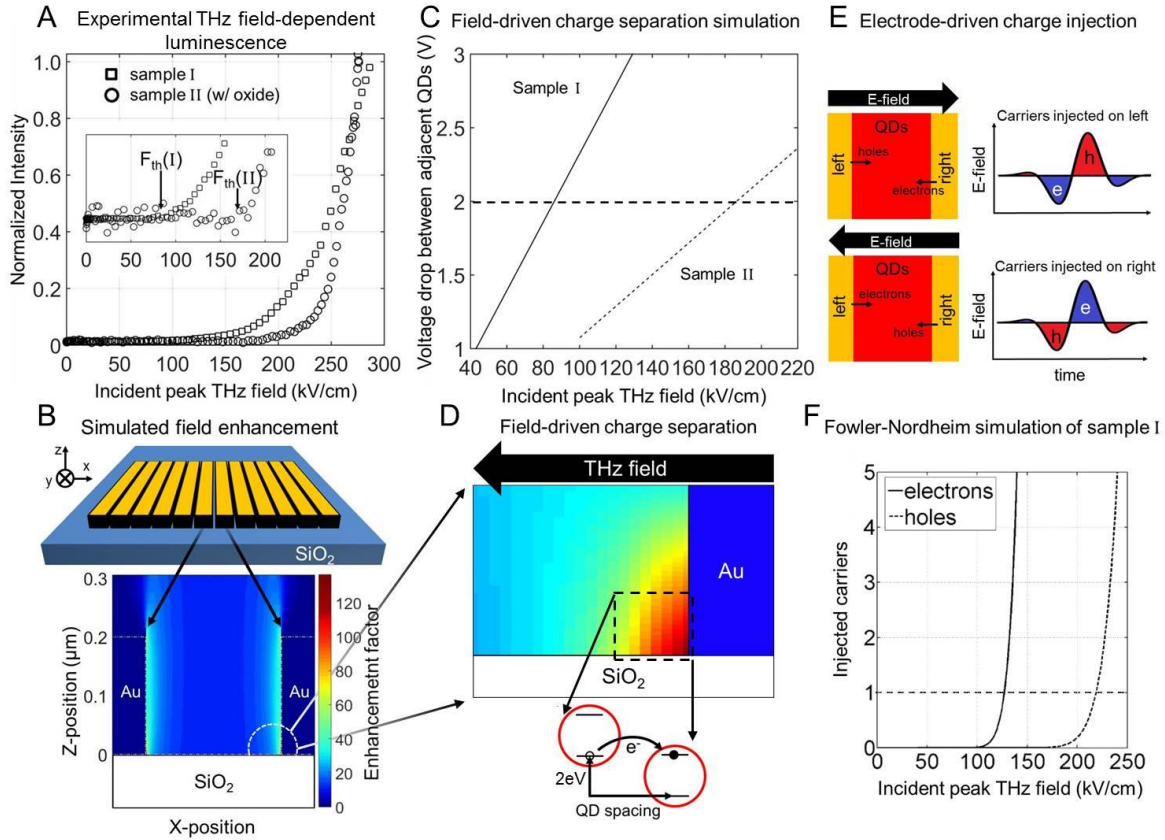

**Fig. 3 Experimental and simulated threshold THz field required to generate luminescence.** (**A**) THz field-dependent luminescence of CdSe:CdS QDs. Varying the incident peak THz field from 0 to 320 kV/cm generates luminescence with a nonlinear dependence. The threshold $F_{th}$ to generate luminescence occurs at 85±5 kV/cm in sample I and 180±7 kV/cm in sample II. (**B**) Simulation of microslit THz near-field enhancement. The enhancement has a spatial variation along the x and z positions and is invariant along y. (**C**) Simulated voltage drop between adjacent QDs as a function of peak incident THz field. The predicted threshold incident THz fields are 86 and 186 kV/cm for sample I and II respectively which is in good agreement with experimental values. (**D**) In the field-driven model, when the voltage drop between QDs is equal to or greater than the QD band gap energy (divided by the electron charge), QD excitons can be generated, leading to luminescence. (**E**) The electrode-driven charge injection model. During the two half-cycles of the THz field, electrons and holes are injected into QDs adjacent to the electrodes, resulting in exciton formation and luminescence. Due to the different THz field profiles in the two polarities, at each electrode more carriers of one type than the other will be injected. (**F**) Using the Fowler-Nordheim model, the number of injected electrons and holes at the inner face of one side of a slit are computed (by symmetry the opposite side will give the same result with electrons and holes reversed). Since exciton generation requires both electrons and holes, the threshold for THz-L will be determined by the higher of the two thresholds for charge injection (holes for one side, electrons for the other). The onset of THz-L is predicted to begin at 219 kV/cm which is much higher than the experimental result for sample I.

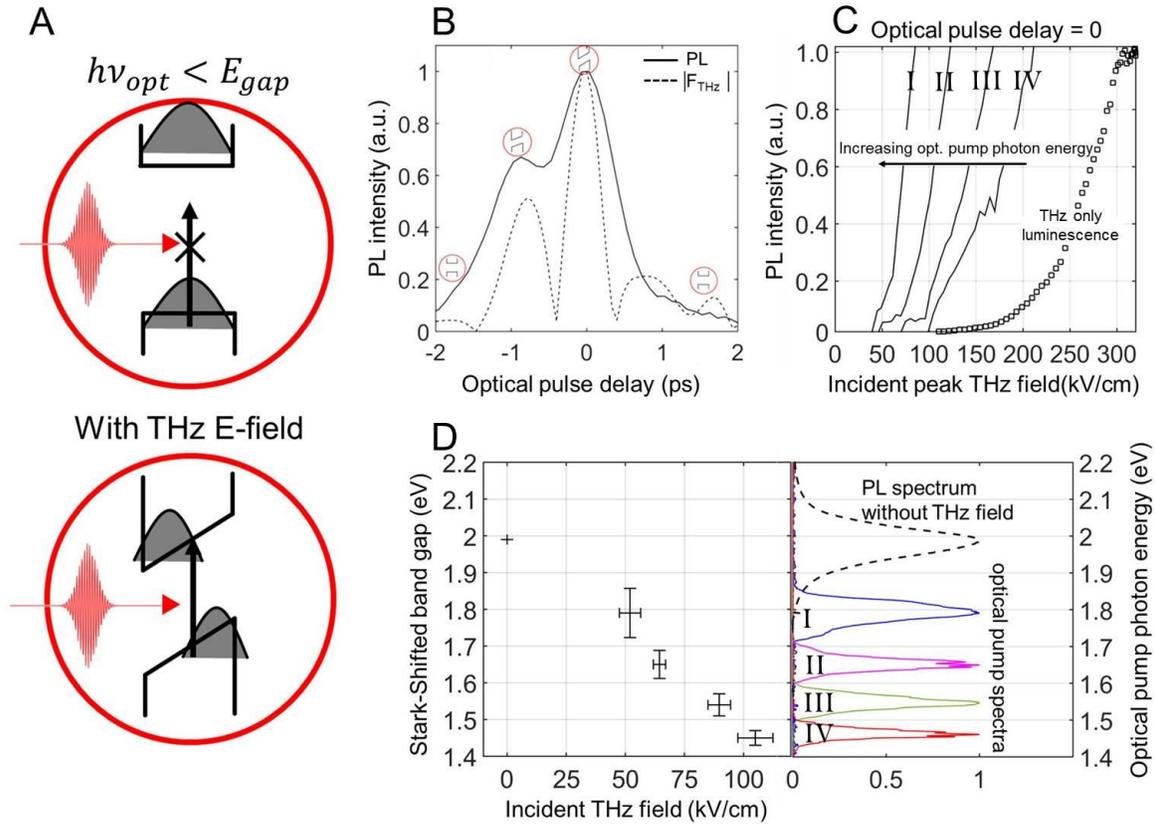

**Fig. 4 THz field-induced Stark shifting of QDs.** (**A**) Below-band-gap photons with energy $h\nu_{opt}$ are not absorbed until a THz field threshold $F_{th}$ is reached which reduces the QD band gap sufficiently, at which point electroabsorption occurs. (**B**) PL intensity as a function of the time delay between the optical pulse and the THz field peak. The PL intensity generated by the below-bandgap optical photons roughly tracks the absolute value of the THz field. (**C**) The optical pulse delay is set to 0 and the peak THz field is varied from 0 to 320 kV/cm. When the pump photon energy (indicated by numerals corresponding to spectra in (**D**)) is increased, the threshold THz field required to generate a PL signal is decreased. At these threshold THz fields the effective band gap of the QDs has been lowered to equal the optical pulse photon energy. (**D**) THz field-induced Stark shifting of QDs. (right) The optical pump pulse spectra with the photon energy plotted on the y-axis. (left) As the THz field strength is increased, optical pulses with lower photon energy are absorbed and generate PL. The points indicate the threshold field levels (from (**C**)) at which the band gap is lowered to an energy equal to the optical photon energy. For a 100 kV/cm incident THz field the band gap is Stark-shifted by > 0.5 eV.

**References:**

1. L. E. Brus, Electron–electron and electron-hole interactions in small semiconductor crystallites: The size dependence of the lowest excited electronic state. *J. Chem. Phys.* **80**, 4403 (1984).

2. S. A. Empedocles, Quantum-Confined Stark Effect in Single CdSe Nanocrystallite Quantum Dots. *Science (80-. ).* **278**, 2114–2117 (1997).

3. A. J. Bennett *et al.*, Giant Stark effect in the emission of single semiconductor quantum dots. *Appl. Phys. Lett.* **97**, 031104 (2010).

4. D. Bozyigit, O. Yarema, V. Wood, Origins of Low Quantum Efficiencies in Quantum Dot LEDs. *Adv. Funct. Mater.* **23**, 3024–3029 (2013).

5. Z.-B. Wang, H.-C. Zhang, J.-Y. Zhang, Quantum-Confined Stark Effect in Ensemble of Colloidal Semiconductor Quantum Dots. *Chinese Phys. Lett.* **27**, 127803 (2010).

6. M. C. Hoffmann, B. S. Monozon, D. Livshits, E. U. Rafailov, D. Turchinovich, Terahertz electro-absorption effect enabling femtosecond all-optical switching in semiconductor quantum dots. *Appl. Phys. Lett.* **97**, 231108 (2010).

7. J. J. Amaral *et al.*, Magnetic field induced quantum dot brightening in liquid crystal synergized magnetic and semiconducting nanoparticle composite assemblies. *Soft Matter*. **11**, 255–260 (2015).

8. M. Liu *et al.*, Terahertz-field-induced insulator-to-metal transition in vanadium dioxide metamaterial. *Nature*. **487**, 345–8 (2012).

9. M. C. Hoffmann, J. Hebling, H. Y. Hwang, K.-L. Yeh, K. A. Nelson, Impact ionization in InSb probed by terahertz pump—terahertz probe spectroscopy. *Phys. Rev. B*. **79**, 161201 (2009).

10. H. Hirori *et al.*, Extraordinary carrier multiplication gated by a picosecond electric field pulse. *Nat. Commun.* **2**, 594 (2011).

11. C. Lange *et al.*, Extremely Nonperturbative Nonlinearities in GaAs Driven by Atomically Strong Terahertz Fields in Gold Metamaterials. *Phys. Rev. Lett.* **113**, 227401 (2014).

12. K. Iwaszczuk, M. Zalkovskij, A. C. Strikwerda, P. U. Jepsen, Nitrogen plasma formation through terahertz-induced ultrafast electron field emission. *Optica*. **2**, 116 (2015).

13. I. Al-Naib *et al.*, Effect of local field enhancement on the nonlinear terahertz response of a silicon-based metamaterial. *Phys. Rev. B*. **88**, 195203 (2013).

14. K. Fan *et al.*, Nonlinear Terahertz Metamaterials via Field-Enhanced Carrier Dynamics in GaAs. *Phys. Rev. Lett.* **110**, 217404 (2013).

15. A. T. Tarekegne, K. Iwaszczuk, M. Zalkovskij, A. C. Strikwerda, P. U. Jepsen, Impact ionization in high resistivity silicon induced by an intense terahertz field enhanced by an antenna array. *New J. Phys.* **17**, 043002 (2015).

16. K. G. Lee, Q.-H. Park, Coupling of Surface Plasmon Polaritons and Light in Metallic Nanoslits. *Phys. Rev. Lett.* **95**, 103902 (2005).

17. M. Shalaby *et al.*, Skirting terahertz waves in a photo-excited nanoslit structure. *Appl. Phys. Lett.* **104**, 171115 (2014).

18. J. M. Caruge, J. E. Halpert, V. Wood, V. Bulović, M. G. Bawendi, Colloidal quantum-dot light-emitting diodes with metal-oxide charge transport layers. *Nat. Photonics*. **2**, 247–250 (2008).

19. D. S. Ginger, N. C. Greenham, Charge injection and transport in films of CdSe nanocrystals. *J. Appl. Phys.* **87**, 1361 (2000).

20. D. Yu, C. Wang, P. Guyot-Sionnest, n-Type conducting CdSe nanocrystal solids. *Science*. **300**, 1277–80 (2003).

21. D. V Talapin, C. B. Murray, PbSe nanocrystal solids for n- and p-channel thin film field-effect transistors. *Science*. **310**, 86–9 (2005).

22. V. Wood, J. E. Halpert, M. J. Panzer, M. G. Bawendi, V. Bulović, Alternating current driven electroluminescence from ZnSe/ZnS:Mn/ZnS nanocrystals. *Nano Lett.* **9**, 2367–71 (2009).

23. W. Mu *et al.*, Direct-Current and Alternating-Current Driving Si Quantum Dots-Based Light Emitting Device. *IEEE J. Sel. Top. Quantum Electron.* **20**, 206–211 (2014).

24. S. H. Cho *et al.*, High performance AC electroluminescence from colloidal quantum dot hybrids. *Adv. Mater.* **24**, 4540–6 (2012).

25. D. Bozyigit, V. Wood, Y. Shirasaki, V. Bulovic, Study of field driven electroluminescence in colloidal quantum dot solids. *J. Appl. Phys.* **111**, 113701 (2012).

26. L. Tavares, J. Kjelstrup-Hansen, H.-G. Rubahn, Localized and guided electroluminescence from roll printed organic nanofibres. *Nanotechnology*. **23**, 425203 (2012).

27. X. Liu *et al.*, AC-biased organic light-emitting field-effect transistors from naphthyl end-capped oligothiophenes. *Org. Electron.* **11**, 1096–1102 (2010).

28. T. Yamao, Y. Shimizu, K. Terasaki, S. Hotta, Organic Light-Emitting Field-Effect Transistors Operated by Alternating-Current Gate Voltages. *Adv. Mater.* **20**, 4109–4112 (2008).

29. V. Wood *et al.*, Electroluminescence from nanoscale materials via field-driven ionization. *Nano Lett.* **11**, 2927–32 (2011).

30. V. Wood *et al.*, Air-stable operation of transparent, colloidal quantum dot based LEDs with a unipolar device architecture. *Nano Lett.* **10**, 24–9 (2010).

31. CST.com (2016), (available at www.cst.com/products/cstmws).

32. P. S. Davids, S. M. Kogan, I. D. Parker, D. L. Smith, Charge injection in organic light-emitting diodes: Tunneling into low mobility materials. *Appl. Phys. Lett.* **69**, 2270 (1996).

33. I. D. Parker, Carrier tunneling and device characteristics in polymer light-emitting diodes. *J. Appl. Phys.* **75**, 1656 (1994).

34. R. H. Fowler, L. Nordheim, Electron Emission in Intense Electric Fields. *Proc. R. Soc. A Math. Phys. Eng. Sci.* **119**, 173–181 (1928).

35. K.-L. Yeh, M. C. Hoffmann, J. Hebling, K. A. Nelson, Generation of 10 μJ ultrashort terahertz pulses by optical rectification. *Appl. Phys. Lett.* **90**, 171121 (2007).

36. P. C. M. Planken, H.-K. Nienhuys, H. J. Bakker, T. Wenckebach, Measurement and calculation of the orientation dependence of terahertz pulse detection in ZnTe. *J. Opt. Soc. Am. B*. **18**, 313 (2001).

37. N. C. J. van der Valk, T. Wenckebach, P. C. M. Planken, Full mathematical description of electro-optic detection in optically isotropic crystals. *J. Opt. Soc. Am. B*. **21**, 622 (2004).

38. †,‡ Luigi Carbone *et al.*, Synthesis and Micrometer-Scale Assembly of Colloidal CdSe/CdS Nanorods Prepared by a Seeded Growth Approach (2007).

39. J. C. de Mello, H. F. Wittmann, R. H. Friend, An improved experimental determination of external photoluminescence quantum efficiency. *Adv. Mater.* **9**, 230–232 (1997).

**Acknowledgments:**

W.C., V.B., J.S., I.C., M.B., and K.A.N were supported as part of the Center for Excitonics, an Energy Frontier Research Center funded by the US Department of Energy, Office of Science, Office of Basic Energy Sciences under Award Number DE-SC0001088. H.Y.H. and B.P. were supported by Office of Naval Research grants no. N00014-13-1-0509 and N00014-16-1-2090.

**Supplementary Materials:**

Materials and Methods

Supplementary Text

Figures S1-S6

References (*35-39*)

**Materials and Methods:**

Experimental apparatus

Vertically polarized THz pulses are generated in $LiNbO_3$ using tilted pulse front phase velocity matching(*35*) and focused onto the samples using a set of off-axis parabolic mirrors (Fig. S1a). The laser amplifier used to generate the THz pulses produces 6 mJ, 800 nm, 35 fs pulses. The time-dependent THz electric field was measured with electro-optic sampling(*36*, *37*) using a 100 µm thick 110-cut gallium phosphide crystal. At the focus, the maximum electric field of the THz pulses is 320 kV/cm (Fig. S1b) with a spectrum centered at 0.5 THz (Fig. S1c). The peak THz electric field is adjusted using a pair of wire-grid polarizers. Ultrafast optical pulses, of select wavelengths between 400nm and 800nm, are generated using a homebuilt non-collinear optical parametric amplifier (NOPA). The average NOPA pulse energy is 50 nJ and the duration ranges from 50-100 fs depending on the wavelength. For the quantum confined Stark effect measurements, the NOPA output is attenuated to prevent PL originating from two-photon absorption. The optical pulses are focused onto the sample with a ~0.75 mm diameter which is slightly smaller than the ~1 mm diameter of the THz focus. Light from the middle of focus was collected to assure THz-L and PL signals originated from the same locations on the sample. Luminescence signals were imaged onto a spectrometer (Andor) or avalanche photodiode (APD, id quantique). The spectrometer has an f/4 aperture, a 100 µm slit and 0.2 nm/pixel dispersion over a 1024x256 back-illuminated CCD camera. Two f/4 lenses were used to image the luminescence 1:1 onto the spectrometer slit or APD. A 650 nm shortpass and 550 nm longpass filter are placed between the collection and imaging lenses to pass only the THz-L and PL signals.

Sample fabrication

Microslit FESs were fabricated using standard lift-off lithography techniques. A negative resist (AZ5214) was spin cast on quartz substrates, UV-exposed through a chrome lithography mask (Karl Suss MA4), and developed (AZ400K) to define the specified gap features and microslit dimensions. A thin film of 15 nm chrome was thermally evaporated on top as an adhesive layer followed by 200 nm of gold. Acetone was used to liftoff the remaining resist features resulting in the gold microslit FES samples. Sample II had an additional 50-80 nm layer of $SiO_2$ thermally evaporated onto the surface before addition of the QD layer. The QDs were spin cast onto the surface of the microslit samples. Figure S2 shows an SEM image of a single microslit gap without oxide coating. The gap width was measured to be 2 µm.

QD synthesis

- Chemicals

1-octadecene (ODE, 90%), trioctylphosphine oxide (TOPO 99%), trioctylphosphine (TOP, 97%), oleylamine (OAm, 70%), 1-octanethiol (> 98.5%), sulphur powder (99.999%) were obtained from Sigma Aldrich. Cadmium oxide (CdO, 99.998%), selenium powder (99.999%), oleic acid (OLA, 90%) and octadecylphosphonic acid (ODPA, 97%) were purchased from Alfa Aesar.

- Synthesis of CdSe Cores

The synthesis of the CdSe cores and the first fast-injection shell growth was carried out by following a previously published.(*38*) The first absorption feature of the cores was at 561 nm.

- Synthesis of the CdSe/CdS Samples

To a 250mL round bottom flask was added 200nmol of the CdSe cores (with a maximum of the first excitonic feature in the absorbance spectrum at 561nm) dissolved in hexane, 5mL of ODE, and 5mL of oleylamine. The solution was degassed at r.t. for 2 hr. and then for 5 min at 100°C to remove the hexane and water. The solution was then stirred under $N_2$ and the temperature was raised to 310°C.

- Injection 1

At 200°C, a solution of Cd-oleate (4mL of a 0.2M solution of Cd-oleate in ODE) dissolved in ODE and a separate solution of octanethiol (166µL) dissolved in ODE (for a total volume of 10mL for each) were injected at a rate of 5.0mL/hr. After the injection stopped after 2hrs, an aliquot of 6mL was taken after waiting for 5 additional minutes.

- Injection 2

Once the aliquot was removed, the same amount of precursors as used in injection 1 was added at a rate of 3.33mL/hr. After the injection stopped after 1.5hrs, a 10mL aliquot was removed after waiting for 5 minutes. Oleic acid (2mL) was then added.

- Injection 3

The same amount of precursors as used in injection 1 was again added at a rate of 3.33mL/hr. After the injection ended after 1.5hrs, the reaction was stopped by removing the heating mantle.

- Purification

Each aliquot was purified via precipitation of the particles using acetone followed by centrifugation at 5000RPM for 5 min. The particles were then re-suspended in hexane.

QD size and PL quantum yield

Figure S3 shows TEM measurements of the QDs used. The CdSe cores where measured to be 4.1 nm in diameter and the additional CdS shell is 2.2 nm thick. From the TEM measurements the inter-dot spacing is estimated to be 10 nm. Figure S4 shows the absorption spectrum of a thin film of QDs on quartz measured for the wavelength range of 400 - 800 nm in 1nm steps using a UV-vis-NIR spectrophotometer (Cary 5000). The PL quantum yield was 19 % measured using a 405 nm laser excitation in an integration sphere setup. A calibrated visible spectrometer and CCD camera setup (Princeton Instrument Acton SP2300 and Pixis) was used to detect wavelength-resolved PL and laser pump intensity.(*39*)

Electromagnetic simulations of microslit field enhancement

Simulations were performed using CST Microwave Studio using the time-domain transient solver.(*31*) Extremely fine mesh sizes were used to understand field localization effects. Near enhancement regions, the mesh was further reduced to single nm sizes to make sure enhancement factors were known to within the length scale of a single quantum dot. Material properties were taken from the CST material library or by assuming constant real and imaginary dielectric or conductivity.

**Supplementary Text:**

Electrode-driven carrier injection simulation

For the experimental THz field $F_{THz}(t)$, measured using electro optic sampling, the current density $J_{FN}$ at the QD-gold interface is

(1) $$J_{FN}(t) = a\left(\frac{F_{THz}(t)}{\varepsilon_{film}}\beta_{gold\;surface}\right)^2 \varphi^{-1} \exp\left(\frac{-b\varepsilon_{film}\varphi^{\frac{3}{2}}}{\beta F_{THz}(t)}\right)$$

where $a = 1.53 \cdot 10^{-6}$ AeVV$^{-2}$ and $b = 6.829 \cdot 10^9$ eV$^{-3/2}$Vm$^{-1}$ are combinations of universal constants. The average (from top free surface to bottom oxide interface) field enhancement across the surface of the gold microslits is $\beta = 106$. The potential energy barrier height $\varphi$ is the barrier electrons or holes must overcome. The height can take on a range of values depending on the nature of the QD/gold interface. We assume that the Fermi level of the gold microslits lies in the middle of the QD band gap, giving an estimated barrier of 1.0 eV for electrons and holes. The THz field $F_{THz}(t)$ has a time-dependent change in polarity that will change which type of carrier, electron or hole, is injected (Fig. S5a). Considering the gold surface on the inside of a single gap at time $t$, electrons will be injected at one gold surface while holes will be injected at the other (Fig. S5b). Focusing on one side of a gap, when the field is positive the number of electrons injected in the field of view of the spectrometer is

(2) $$N_e = A_{slit} \sum_i J_{FN}(t_i, F_{THz}(t_i)) \cdot \Delta t \quad , \text{for positive } F_{THz}$$

and when the field is negative the number of holes injected is

(3) $$N_h = A_{slit} \sum_i J_{FN}(t_i, F_{THz}(t_i)) \cdot \Delta t \quad , \text{for negative } F_{THz}$$

where $A_{slit}$ is the cross-sectional area of the slit that is imaged into the spectrometer and $\Delta t$ is the time step size. The gap walls are 200 nm tall and the spectrometer images a 100 µm wide portion of each side of the slit, making $A_{slit} = (200\text{nm}) \cdot (100\text{µm})$. The summation is over the number of time steps used to experimentally measure $F_{THz}$(t). Here we have made the assumption that changing the sign does not inject the carriers back into the gold surface. Any such return to the gold would reduce the luminescence observed, potentially raising the threshold THz field level for observable luminescence signal even farther above the level measured experimentally.

PL intensity dependence on optical excitation pump

In the electroabsorption experiments, THz induced two-photon absorption is a possibility that could complicate our interpretation. To rule this out, the optical power was varied from 0 to 400 µW at zero time delay between the optical pulse and the THz field peak, and the PL intensity was monitored (Fig. S5). We show results from 800 nm pump experiments which are representative of the results at shorter and longer optical pump wavelengths. The PL follows a linear dependence on the incident optical power making it clear that the THz field modulated the linear absorption cross section. It should be noted that without THz pulses there was an observable PL signal at the highest optical power which originates from the two-photon absorption cross section already present in the QDs; this component has been subtracted from the optical power-dependent data in Figure S6.

**Supplementary Materials Figures:**

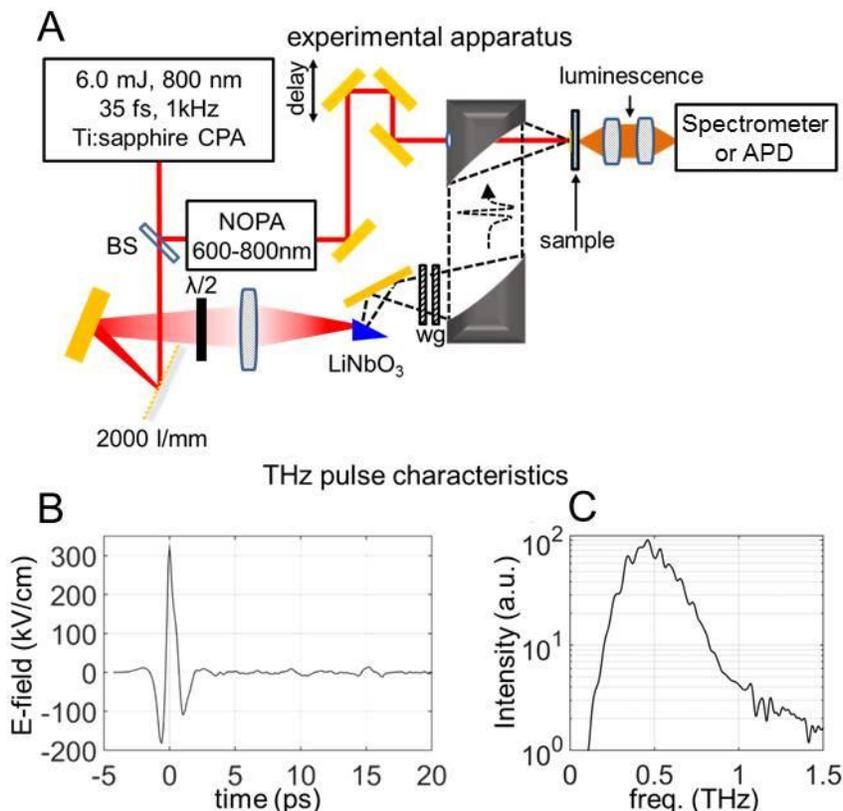

**Fig. S1 Experimental apparatus and THz pulse characteristics.** (**A**) THz pulses are generated using the tilted pulse front phase velocity matching method and focused using a pair of off-axis parabolic mirrors. A pair of wire-grid (wg) polarizers varies the incident THz E-field onto the sample. A spectrometer measures the visible light emission. Using a beam splitter (BS), a portion of the 800 nm laser pulse was redirected to pump the NOPA, producing tunable 600 nm - 800 nm pulses for sub-band gap optical excitation of QDs in electroabsorption experiments. (**B**) The THz pulses have a maximum 320 kV/cm electric field. (**C**) The spectrum of the THz pulses is centered at 0.5 THz.

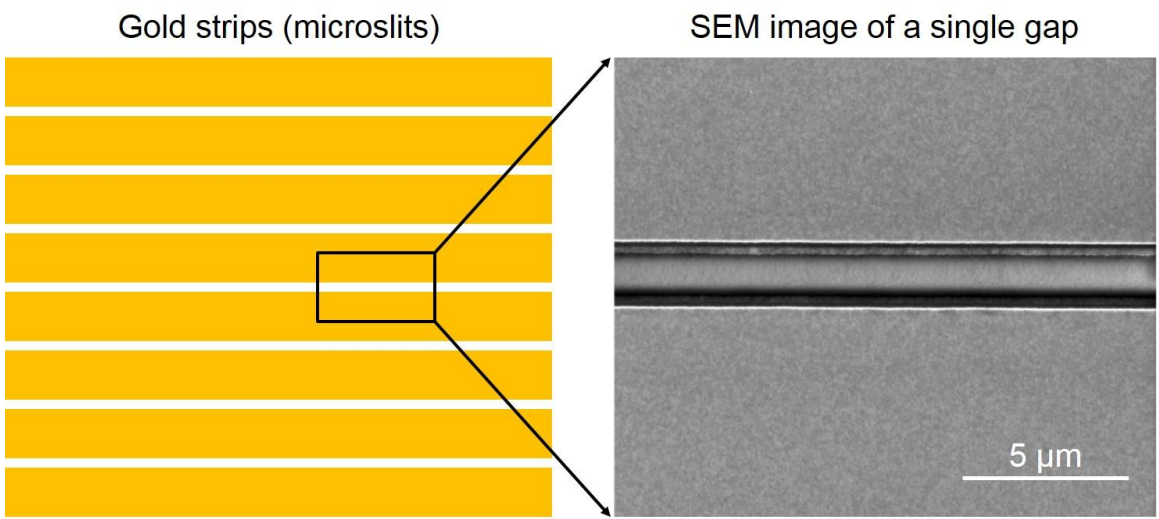

**Fig. S2 SEM image of microslits.** The gold strips are separated by 2 μm wide capacitive gaps.

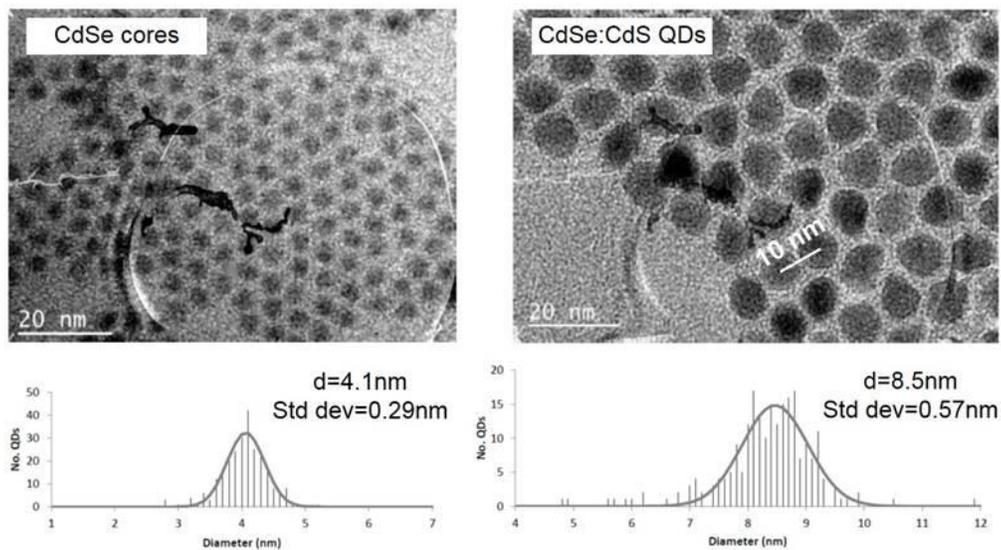

**Fig. S3 SEM images of quantum dots.** The CdSe cores have an average diameter of 4.1 nm. The addition of the CdS shell increases the average diameter to 8.5 nm, corresponding to a 2.2 nm thick shell. Accounting for the space between the QDs due to the passivation ligands, the inter-dot spacing is ~10 nm.

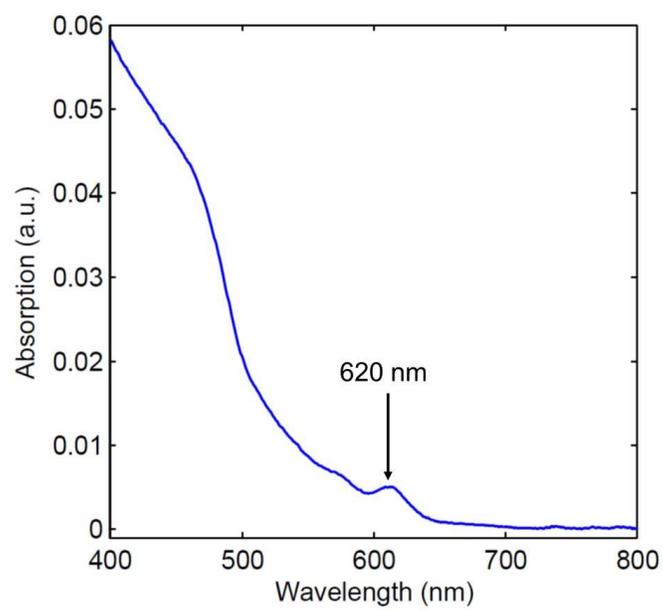

**Fig. S4 Absorption spectrum of QD thin film.** The absorption was measured using a UV/vis spectrometer. The measured sample was a thin film of QDs spin-cast onto a quartz substrate. The exciton absorption is clearly visible at 620 nm.

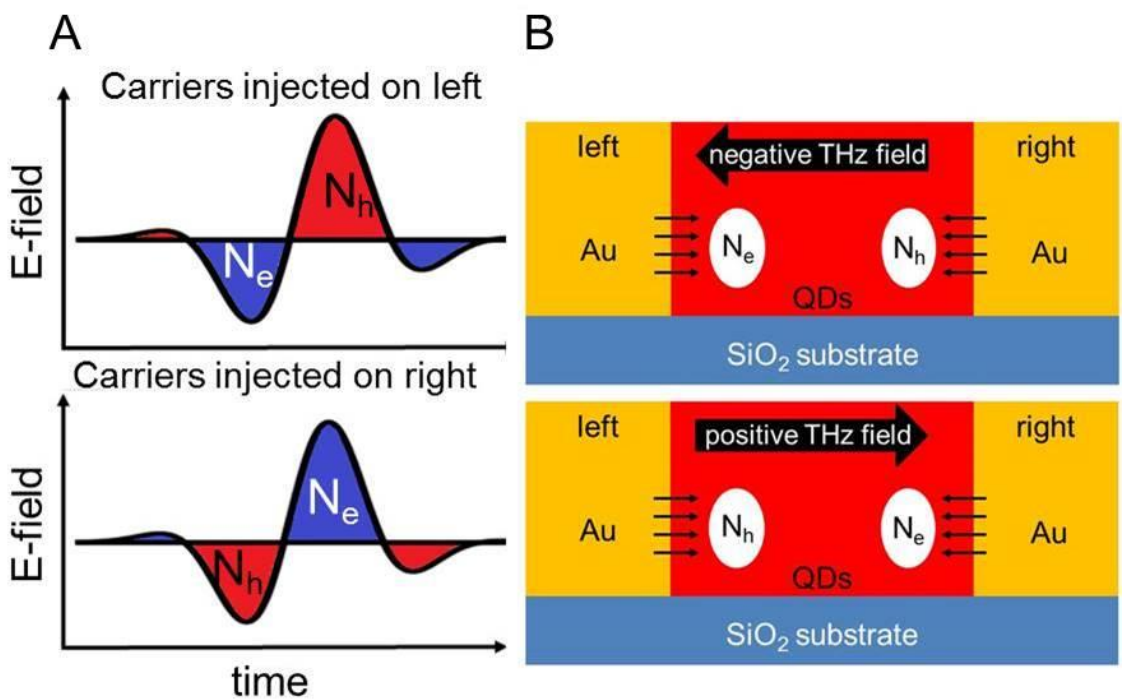

**Fig. S5 Electrode-driven carrier injection modeling.** (**A**)The time-dependent THz field has positive and negative field components that inject electrons or holes depending on the side of the gap considered. (**B**) For a negative (left pointing) field polarity electrons are injected from the left side of a gap and holes from the right. For a positive (right pointing) field polarity holes are injected on the left and electrons from the right.

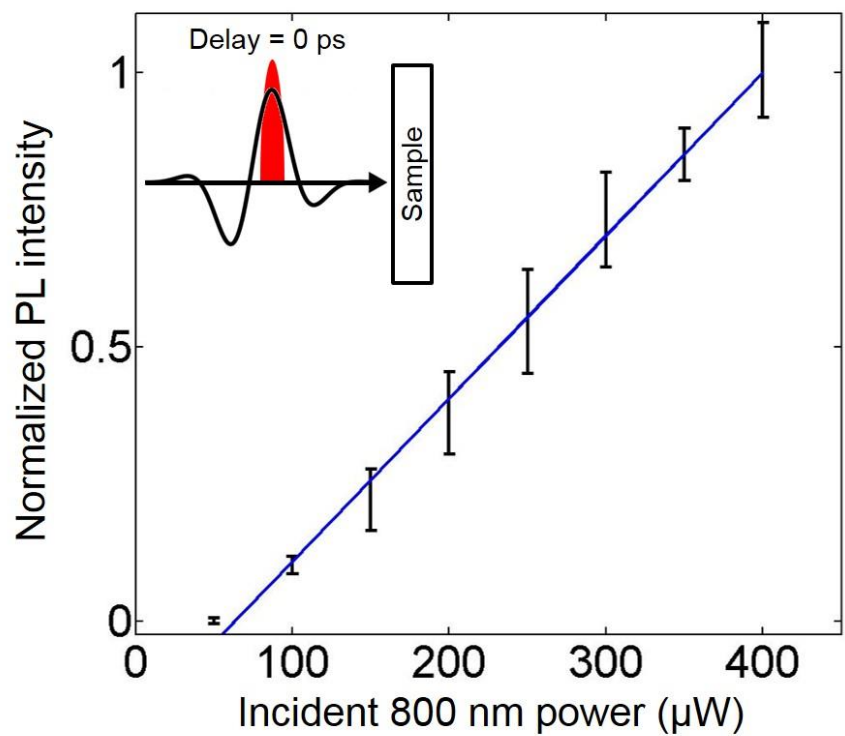

**Fig. S6 Dependence of electroabsorption PL on incident optical power.** To rule out the possibility of THz-induced two-photon absorption, the optical power was varied at zero THz pump-optical pump delay (inset). The PL signal originating from electroabsorption has a clear linear dependence. Significant two-photon absorption would generate quadratic components in the power dependence which are not apparent here. Therefor our THz-driven electroabsorption measures change in the linear absorption spectrum.